\documentclass[prd,aps,twocolumn,a4paper,showkeys,nofootinbib]{revtex4-1}

\usepackage{graphicx,psfrag}
\usepackage{mathrsfs}
\usepackage{amsmath,amsfonts,amssymb}
\usepackage{multirow}
\usepackage{comment}
\usepackage{hyperref}
\usepackage{enumitem}
\usepackage{tcolorbox}

\newcommand{\be}{\begin{equation}}
\newcommand{\ee}{\end{equation}}
\newcommand{\bea}{\begin{eqnarray}}
\newcommand{\eea}{\end{eqnarray}}
\newcommand{\bel}{\begin{align}}
\newcommand{\eel}{\end{align}}

\def\l{\ell}
\def\lm{{\ell m}}
\def\p{\partial}

\def\non{\nonumber}

\def\i{{\rm i}}

\def\Msun{{\rm M_{\odot}}}
\def\Mo{\Msun}

\def\GMc2{{\rm G M_{\odot} c^{-2}}}

\def\O{\mathcal{O}}

\def\scri{{\mathcal{I}^+}}
\def\ie{{\it i.e.}}
\def\eg{{\it e.g.}}
\def\cf{{\it cf.}}

\newcommand{\rwzhyp}{\texttt{RWZHyp}}
\newcommand{\gra}{\texttt{GR-Athena++}}
\newcommand{\pittnull}{\texttt{PittNull}}
\newcommand{\spectre}{\texttt{SpECTRE}}

\usepackage{pifont} 

\usepackage{color}
\definecolor{cyan}{rgb}{0,0.9,0.9}
\definecolor{orange}{rgb}{0.9,0.5,0}
\definecolor{magenta}{rgb}{1,0,1}
\definecolor{purple}{rgb}{0.8,0.4,0.8}
\definecolor{gray}{rgb}{0.8242,0.8242,0.8242}
\definecolor{light-gray}{gray}{0.95}

\begin{document}

\title{Perturbative Hyperboloidal Extraction\\
  of Gravitational Waves in 3+1 Numerical Relativity}

\author{Sebastiano \surname{Bernuzzi}$^1$}
\author{Joan \surname{Fontbuté}$^1$}
\author{Simone \surname{Albanesi}$^{1,2}$}
\author{An{\i}l \surname{Zengino\u{g}lu}$^3$}
\affiliation{$^1$~Theoretisch-Physikalisches Institut, Friedrich-Schiller-Universit{\"a}t Jena, 07743, Jena, Germany}
\affiliation{$^2$~INFN Sezione di Torino, Via P. Giuria 1, 10125 Torino, Italy}
\affiliation{$^3$~Institute for Physical Science and Technology, University of Maryland, College Park, MD 20742, USA}

\date{\today}

\begin{abstract}
  We present a framework to propagate to null infinity gravitational
  waves computed at timelike worldtubes in the interior of a  
  3+1 (Cauchy) numerical relativity simulations.
  In our method, numerical relativity data are used as the inner inflowing boundary of a
  perturbative time-domain Regge-Wheeler-Zerilli simulation in
  hyperboloidal coordinates that reaches null infinity.
  We showcase waveforms from (3+1)D simulations of
  gravitational collapse of rotating neutron stars, binary black holes
  mergers and scattering, and binary neutron star mergers and compare
  them to other extrapolation methods. 
  Our perturbative hyperboloidal extraction provides a simple yet
  efficient procedure to compute gravitational waves with data quality comparable to the 
  Cauchy characteristic extraction for several practical applications.
  Nonlinear effects in the wave propagation are not captured by our method, but 
  the present work is a stepping stone towards more
  sophisticated hyperboloidal schemes for gravitational-wave extraction. 
\end{abstract}

\maketitle

\section{Introduction}
\label{sec:intro}

Gravitational-wave astronomy relies on numerical relativity (NR)
computations for accurate predictions of gravitational waves from
strongly gravitating and fast moving sources. Examples of such sources
are stars undergoing gravitational collapse and compact binary mergers.
Formalisms and algorithms for wave extraction play a key role for the computation of NR waveforms; the main target is to compute unambiguous 
data at future null infinity ($\scri$).

In the context of 3+1 NR with Cauchy-type foliations, waveforms can be
extracted at coordinate spheres at finite radii on spacelike hypersurfaces.
Waveform extraction algorithms are build on either on the
Regge-Wheeler-Zerilli-Moncrief (metric) formalism for perturbations of spherical
spacetimes~\cite{Abrahams:1995gn,Abrahams:1997ut,Rupright:1998uw,Camarda:1998wf,Allen:1998wy,Pazos:2009vb,Fontbute:2025ixd}
or the Newmann-Penrose (curvature) formalism for perturbations of
Kerr-spacetimes (in particular using Weyl's $\psi_{4}$ scalar) \cite{Newman:1966}.
Independently on the specific technique, these waveforms are affected
by two main systematic errors:
i) the ``artificial'' timelike boundary conditions of the main
evolution system, 
and
ii) the finite radius of extraction. 
A Cauchy perturbative matching extraction algorithm,
in which the Cauchy evolution of the strong-field spacetime region is matched
to a perturbative evolution that provides both waveforms 
and boundary condition was proposed to alleviate these
issues~\cite{Abrahams:1997ut,Rupright:1998uw}. However, the method
still suffers of finite extraction error and it has been abandoned in
modern simulations.

A commonly employed technique for the computation of waveforms at $\scri$ is to
extrapolate in radius and (approximate) retarded time,
\eg~\cite{Scheel:2008rj,Lousto:2010qx,Bernuzzi:2011aq,Nakano:2015pta}.  
Alternatively, a more sophisticated approach is the Cauchy-characteristic extraction
(CCE). In CCE, data from a worldtube in the interior of a Cauchy domain
is propagated to null infinity using a null-cones formulation of Einstein
equations, \ie~a characteristic
evolution~\cite{Bishop:1996gt,Bishop:1997ik,Babiuc:2008qy,Reisswig:2006nt,Reisswig:2009us,Reisswig:2009rx,Moxon:2020gha}.  
CCE allows, in principle, a more rigorous extraction of waveforms at
$\scri$. However, it retains the boundary condition issue and it
is, in practice, sensitive to the choice of the worldtube~\cite{Reisswig:2009rx,Moxon:2020gha,Rashti:2024yoc}. 
The development of CCE was originally motivated as a preparatory
step for Cauchy-characteristic
matching~\cite{dInverno:1996tcu,Bishop:1996gt,Bishop:1998uk}. In the latter approach,  
the Cauchy evolution of the strong-field spacetime region is matched
to a characteristic evolution that provides both waveforms at $\scri$
and boundary conditions, see \cite{Winicour:2008vpn} for a review.
Significant progress has recently been made with this technique,
\eg~\cite{Bishop:1998uk,Reisswig:2009us,Moxon:2020gha,Ma:2023qjn,Ma:2024hzq},
which can deliver robust waveforms for modeling purposes.
Note however that the characteristic formulation of the Einstein field
equations is weakly hyperbolic with potential implications on the
convergence of numerical 
data~\cite{Giannakopoulos:2023zzm}.

The use of hyperboloidal foliations in applications to gravitational-wave
astronomy and waveform extraction is less developed.
Hyperboloidal foliations provide a natural framework for the initial-boundary value
problem in General Relativity and has been extensively studied in
mathematical relativity, \eg~\cite{Friedrich:1983,Friedrich:1986,Frauendiener:1997zc,Frauendiener:1997ze,Frauendiener:1998yi,Zenginoglu:2008pw} (and \cite{Frauendiener:2000mk,Frauendiener:2002mm,PanossoMacedo:2023qzp,Zenginoglu:2025sft} for reviews). 
However, the hyperboloidal approach is not yet fully
established for astrophysical solutions to the full
Einstein equations in (3+1)D.
Recent efforts included a full solution of the perturbation problem in
hyperboloidal coordinates~\cite{Zenginoglu:2007jw,Zenginoglu:2009ey,Zenginoglu:2010cq,Bernuzzi:2011aj}, advances in the understanding of the quasinormal mode spectrum \cite{Jaramillo:2020tuu,Gasperin:2021kfv,Cao:2024oud,PanossoMacedo:2024nkw,DeAmicis:2025xuh},
spherically symmetric black-hole evolutions with $\scri$-fixing compactified hyperboloidal coordinates~\cite{Vano-Vinuales:2014koa,Bhattacharyya:2021dti,Vano-Vinuales:2024tat,Peterson:2024bxk}, 
dual foliation approach with asymptotic
regularization~\cite{Hilditch:2016xzh,Gasperin:2019rjg,Duarte:2022vxn}
and the first simulation of gravitational waves from past to future null-infinity with conformal field
equations~\cite{Frauendiener:2025xcj}. 

This work puts forward a simple and yet effective hyperboloidal method
to propagate to null infinity the gravitational waveforms computed in a 3+1 NR
simulation. The method builds on the CCE idea of
propagating data from a worldtube in the interior of a Cauchy domain
to null infinity. However, here the propagation is performed
perturbatively using compactified hyperboloidal slices of
Schwarzschild spacetime. 
This \emph{perturbative hyperboloidal extraction} (PHE) is described 
in Sec.~\ref{sec:meth}. 

The PHE method is validated in Sec.~\ref{sec:res}, where PHE waveform
at $\scri$ are presented from (3+1)D simulations of gravitational
collapse, binary black holes and binary neutron stars mergers.
PHE waveforms are systematically compared to waveforms extracted at
finite-radius, waveform extrapolated in radius and CCE waveforms 
computed from the same simulations.

The paper ends with conclusions which discuss limitations and future
development towards nonlinear evolutions with 
hyperboloidal foliations.

Throughout this paper we use geometric unit, $G=c=1$ and express
masses and lengths in solar masses ($\Mo$). Partial
derivatives are indicated with a suffix, e.g. $\Psi_x\equiv\p_x\Psi$. 
Complex waveform modes are decomposed into amplitude and phase
according to $\Psi_\lm=A_\lm e^{-i\phi_\lm}$.

\section{Method}
\label{sec:meth}

Numerical relativity simulations in the 3+1 Cauchy approach provide 
waveforms on a timelike worldtube in the interior of the Cauchy 
domain by stacking data in time at a coordinate Schwarzschild
radius $\bar{r}$. The (approximation to the) strain at finite radius
is given by the spin-weighted ($s=-2$) mode expansion
\be
h_+-ih_\times =
\sum_{\ell\geq2}\sum_{\ell=-m}^m
\frac{N_\ell}{D_L}
\left(\Psi^{(e)}_\lm(u)+i\Psi^{(o)}_\lm(u)\right){}^{-2}
Y_{\lm}(\theta,\phi)\ ,
\ee
where $D_L$ is the luminosity distance, $\Psi^{(e/o)}_\lm(u)$ are the modes of
the Regge-Wheeler-Zerilli (RWZ) 
master functions of even and odd parity as a function of the retarded time $u$ and
$N_\ell=\sqrt{(\l+2)!/(\l-2)!}$. In perturbation theory of spherical
spacetimes and far 
from the source, these functions satisfy the 1+1 RWZ equation, 
\be\label{eq:rwz}
\Psi_{tt} - \Psi_{r_*r_*} + V\Psi = 0 \ ,
\ee
which is written above in tortoise coordinates $(t,r_*)$ and where 
$V$ is the $\ell$-dependent potential of the background metric
(superscripts (e/o) and multipolar labels are dropped in the notation   
for $\Psi$ and $V$.)

\begin{figure}[t]
  \centering 
    \includegraphics[width=0.5\textwidth]{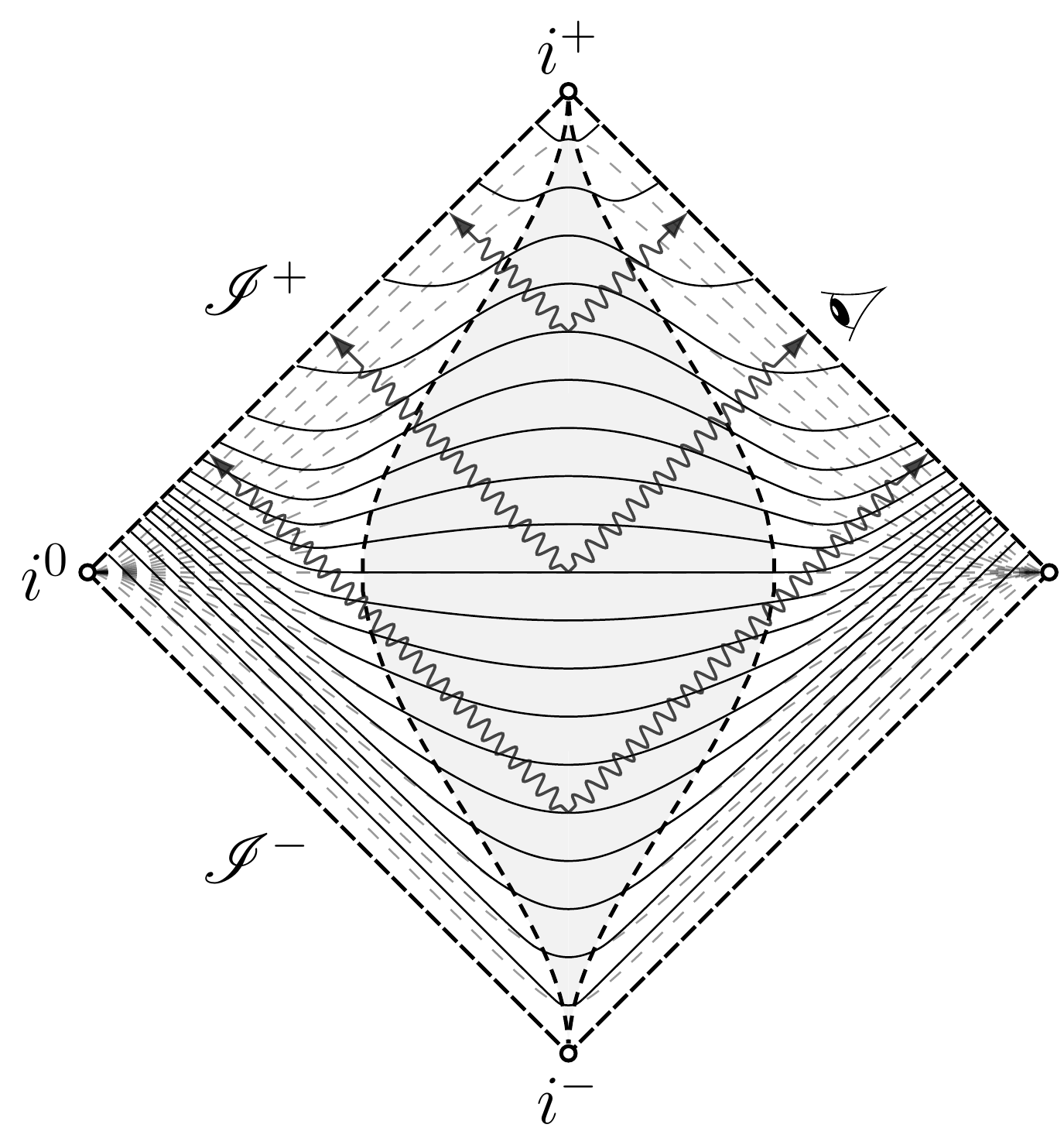}
    \caption{Hyperboloidal layer foliation in a representative Penrose
      diagram of an asymptotically simple spacetime. The shaded region
      contains a truncated Cauchy foliation. 
      Solid lines show the Schwarzschild
      foliation employed in our work~\cite{Bernuzzi:2011aj};
      the two dashed lines connecting past ($i^-$) and future $(i^+)$ 
      timelike infinity indicate the transition layer between Cauchy and
      hyperboloidal coordinates ($r_*=\rho=r_*^{\rm L}$).
      PHE gets inner boundary data at $r_*\lesssim r_*^{\rm L}$ from a
      3+1 NR simulation and propagates them to $\scri$.}
 \label{fig:layer_penrose}
\end{figure}

To perform PHE, we inject the RWZ mode $\Psi^{\rm NR}(t,r)$ 
from a 3+1 NR simulation as ingoing boundary data 
into a 1+1 RWZ evolution code using hyperboloidal slices, 
which propagates the waveform to future null infinity ($\scri$).
Specifically here we employ a hyperboloidal
layer~\cite{Zenginoglu:2010cq, Bernuzzi:2011aj}, which is illustrated
in Fig.~\ref{fig:layer_penrose}. The algorithm and implementation details
are described below. 

The boundary condition for the 1+1 RWZ evolution is prescribed
through the characteristic flowing into the computational domain, 
\be\label{eq:bc:g}
\Psi_t - \Psi_{r_*} = g(t) \ \ \ \mbox{at } \bar{r}_*=r_*(\bar{r}) \ ,
\ee
where $g(t)$ is the given data at the extraction radius $\bar{r}$. The RWZ
equation \eqref{eq:rwz} is then solved on a ``wave-zone'' domain with
$r_*\geq \bar{r}_*=r_*(\bar{r})$.

The time-domain solution at $\scri$ is obtained using the
hyperboloidal layer technique introduced by
\citet{Zenginoglu:2010cq,Zenginoglu:2011jz}. 
We use the setup developed in \cite{Bernuzzi:2011aj}, to which we
refer for all the details.
Hyperboloidal coordinates $(\tau,\rho)$ are obtained from $(t,r_*)$ by
the transformation 
\be
\tau = t - h(r_*) \ ,  \ \
r_* = \frac{\rho}{\Omega(\rho)} \ ,
\ee
where $h(r_*)$ is the height function and the function $\Omega(\rho)$
implements a compactification of the unbounded domain $r_*\in[r_*^{\rm
    L},+\infty)$ to 
  $\rho\in[\rho_*,\rho_S]$. The coordinate $\rho_*=r_*^{\rm 
    L}>\bar{r}_*$ locates the 
  position of the layer, while $\rho_S$ is the coordinate location of $\scri$.
  The time coordinate $\tau$ is fixed by
  demanding that outgoing null rays are invariant in the layer, \ie
\be
u = t - r_* = \tau-\rho 
\ee
which implies $d\rho/dr_*=1-h'(r_*)=:1-H(\rho)$, where $H(\rho)$ is called the
boost function. The transformation is completed with a choice of 
$\Omega(\rho)$ such that $\Omega=1$ for all $\rho<\rho_*$ and some
differentiability across the layer is guaranteed \cite{Bernuzzi:2011aj}.

The RWZ equation on the hyperboloidal slices is 
\begin{align}\label{eq:rwz:hyp}
  -(1+H)\Psi_{\tau\tau} &- 2H \Psi_{\tau\rho} 
  -H_\rho(\Psi_\tau+\Psi_\rho)\, + \\
  &
  + (1-H)\Psi_{\rho\rho} + \frac{V}{1-H}\Psi = 0 \non\ ,
\end{align}
and it reduces to Eq.~\eqref{eq:rwz} for $\rho\leq\rho_*$.
The equation requires no boundary condition at
$\scri$. Equation~\eqref{eq:bc:g} is imposed at the timelike inner boundary
located at the same Schwarzschild radius of the 3+1 extraction sphere,
$\bar{r}_*<r_*^{\rm L}$. 

The numerical implementation follows closely the \rwzhyp{} code
\cite{Bernuzzi:2011aj}. The RWZ 
equation for each mode is solved in first-order in  time and
second-order in space form adopting the method of lines and the
Runge-Kutta 4th order scheme. The right hand side is discretized with
a evenly space grid of $N$ nodes in $\rho\in X=[\rho_{\rm min},\rho_S]_{\rho_*}$, where 
$\rho_{\rm min}=\bar{r}_*$ and $\rho_*=r_*^L>\bar{r}_*$ denotes the interface to the hyperboloidal layer.
Fourth order finite differences are
employed for the derivatives: centered stencils are used in the bulk of
the domain, lop-sided or sided stencils for the outermost points.
No boundary data nor ghost points are used around $\scri$. The inner
boundary data with Eq.~\eqref{eq:bc:g} is
implemented following the 4th order prescription of
\citet{Calabrese:2005fp};
the discrete initial-boundary value problem is well posed in standard energy norms.
Initial data for $\Psi$ and $\Psi_t$ are zero.

The 3+1 NR data employed here were presented in recent work by some of
us \cite{Fontbute:2025ixd}.
The simulations are performed with \gra{}
\cite{Daszuta:2021ecf,Cook:2023bag} using the Z4c free-evolution scheme and the puncture gauge
\cite{Bernuzzi:2009yu,Hilditch:2012fp} and general relativistic
hydrodynamics. We refer the reader to the above
papers for all the details.
Gravitational waves are extracted
at spheres of coordinate (isotropic) radii in the range $R\sim[100,800]$ using a
covariant and gauge invariant 
RWZ metric extraction algorithm that provides $\Psi^{\rm NR}$ at the
extraction sphere \cite{Fontbute:2025ixd}. This choice of
worldtubes is typical for astrophysical simulations; it strikes a 
balance between large extraction radii and the
resolution of the (de-refined) 3D mesh in the wave zone. 

The PHE of these NR data is computed with a grid
$X=[\rho_{\rm
    min},\rho_S]_{\rho_*}=[\bar{r}_*,2\bar{r}_*]_{3\bar{r}_*/2}$
where the innermost point is located at the extraction sphere 
$\rho_{\min}=\bar{r}_*=r_*(\bar{r}(R))$,
the layer is located at $\rho_*=3\bar{r}_*/2$ and
the outermost point ($\scri$) is located at $\rho_S=2\bar{r}_*$.
The resolutions employed in this work are $N=101,201,401,801,1601,3200$.
Simulations with $N\gtrsim1601$ are well converged and truncation errors are pratically negligible in the comparison with other extraction methods.
The RWZ waveforms are also extrapolated in $1/R$; we refer to these waveforms to as ``$R$-extrapolated''. \citet{Fontbute:2025ixd} experimented with different extrapolation formulas and comparison with CCE data. Best results are obtained using a variant of Eq.~(4) in \citet{Nakano:2015pta}, 
\be
\Psi(u) = \left(1-\frac{2M}{\bar{r}}\right)\left(\Psi(u,\bar{r})-\frac{\ell(\ell+1)}{2\bar{r}}\int_0^u dt\,\Psi(t,\bar{r})\right)\,,
\ee
which analytically accounts for $\O(1/r^2)$ terms.
CCE waveforms are computed from the same simulations using the \pittnull~code \cite{Bishop:1998uk,Babiuc:2010ze}. Note that these CCE data are in terms of Weyl's scalar or News
function, and thus require integration to compute the RWZ multipoles
(or equivalently the strain.) Therefore, CCE waveforms have been further
crossed-checked with the \spectre~CCE implementation
\cite{Moxon:2020gha,Ma:2023qjn}, which additionally provides the
strain. We find good agreement between the two output and show the
former data for most of the cases.

\section{Results}
\label{sec:res}

\subsection{Toy boundary data}

\begin{figure}[t]
  \centering 
    \includegraphics[width=0.49\textwidth]{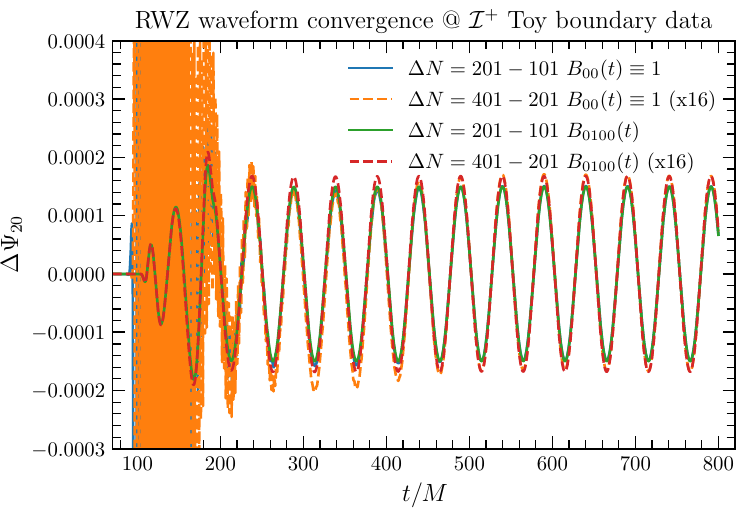}
    \caption{Convergence of RWZ waveforms at $\scri$ with the toy boundary data.
      Differences between pair of waveforms at resolutions
      $N=101,201,401$ with and without using the smooth bump function in
      the boundary data.
      The difference between data at the two highest resolutions are
      rescaled by a factor $16$, corresponding to fourth order
      convergence (dashed lines).} 
 \label{fig:toybd}
\end{figure}

As a first toy problem we consider a Schwarzschild spacetime with mass
$M=1$ and the boundary function
\be
g(t) = B_{ab}(t) \sin(t/8) \ ,
\ee
where $B_{ab}(t)$ is a smooth bump (or activation) function
between $t=a$ and $t=b$ such that $B_{ab}(t\leq a)=0$ and 
$B_{ab}(t\geq b)=1$.
The grid is $X=[100,200]_{150}$ in units of $M$ and the resolutions 
employed are $N=101,201,401$. 

Figure~\ref{fig:toybd} shows the convergence of the RWZ waveform at
$\scri$. The differences between waveforms computed at different
resolutions overlap once rescaled by the appropriate convergence factor
for 4th order accuracy. Some noise is introduced in the convergence
plot if the activation function is not used ($B_{00}(t)\equiv1$). This is
due to the non-smooth injection of boundary data.
The issue can be cured by using the activation function, for example
with parameters $(a,b)=(0,100)$. In the next applications the bump
function is not used since the inner boundary data is sufficiently
smooth and does not introduce noise.

\subsection{Test-mass effective-one-body insplunge}

\begin{figure}[t]
  \centering 
    \includegraphics[width=0.49\textwidth]{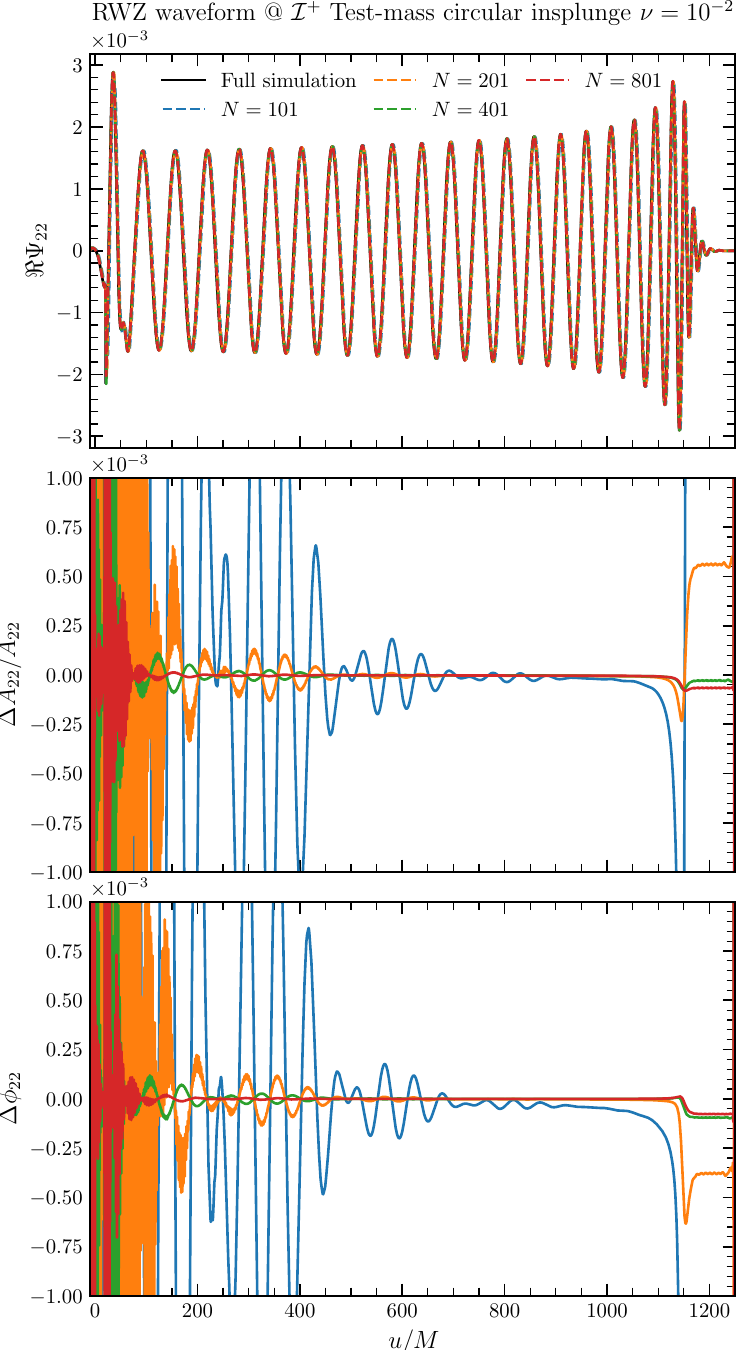}
    \caption{RWZ waveform at $\scri$ from a test-mass insplunge
      simulation.
      Top: Real part of $\Psi_{22}$ from the original simulation
      (black solid line) and from PHE using data t $r_*=100$ and increasing
      resolutions $N=101,201,401,801$.
      Middle: Relative amplitude differences to the full simulation
      for the different resolutions.
      Bottom: Phase differences to the full simulation
      for the different resolutions.}
 \label{fig:testmass}
\end{figure}

The next test performed to validate the PHE method employs the
\rwzhyp{} code to generate a waveform from the inspiral and plunge of
a test-mass source subject to an effective-one-body radiation reaction,
e.g.~\cite{Bernuzzi:2011aj,Bernuzzi:2010xj,Damour:2012ky}. The
radiation force is triggered by setting a symmetric mass ratio of
$\nu=10^{-2}$; the background spacetime mass is $M=1$. 
RWZ waveforms are extracted both in the Cauchy region of the
computational domain, specifically at $\bar{r}_*=100$, and at $\scri$.
The waveform extracted at $\bar{r}_*=100$ is injected in a new set of
PHE simulations~\footnote{The PHE simulations are performed with an
independent code, implementing the same methods.} as boundary data, 
\be
g(t) = \Psi_t^{\rm RWZ} - \Psi_{r_*}^{\rm RWZ} \ .
\ee
The waveform propagates (again) to $\scri$ and is compared to the
result of the full \rwzhyp{} simulation (reference data). This is a
self-consistency check of the method. 
The PHE grid is resolved with $N=101,201,401,801$ points which are at least a factor ten lower
than the reference \rwzhyp{} data (ther latter uses $X=[-50,200]_{150}$ and
$N=12501$.)

Figure~\ref{fig:testmass} shows the 
$\lm=22$ PHE waveform at $\scri$ from various resolution compared to
the reference data. The waveforms are correctly converging to the
reference data with phase differences of
$\Delta\phi_{22}\lesssim10^{-4}$ and relative amplitude differences
$\Delta A_{22}/A_{22}\sim2\times10^{-4}$. 
Note that the phase differences between the $(2,2)$ waveforms
extracted at $r_*=100$ and $\scri$ are larger
than $\Delta\phi_{22}\gtrsim0.1$~rad and out of the plot scale (see
\eg~Fig.~9 of \cite{Bernuzzi:2011aj}.) Similarly, typical amplitude
errors due to finite extraction are $\Delta
A_{22}/A_{22}\gtrsim2\times10^{-3}$.  

\subsection{Gravitational collapse}

Moving to applications of PHE to (3+1)D NR data, we consider waveforms
from the gravitational collapse of an unstable rotating neutron star with mass
$M\simeq1.86$ and uniform roation close to the mass-shedding (Kepler) limit \cite{Fontbute:2025ixd}, 
\cf~\cite{Reisswig:2012nc,Dietrich:2014wja,Cook:2023bag} for similar simulations.
The waveforms $\Psi^{\rm NR}(t)$ extracted at finite radius are
injected in a set of PHE simulations using
\be\label{eq:gNR}
g(t) = \Psi_t^{\rm NR} - \Psi_{r_*}^{\rm NR}\ ,
\ee
and propagated to $\scri$. 

\begin{figure}[t]
  \centering
  \includegraphics[width=.49\textwidth]{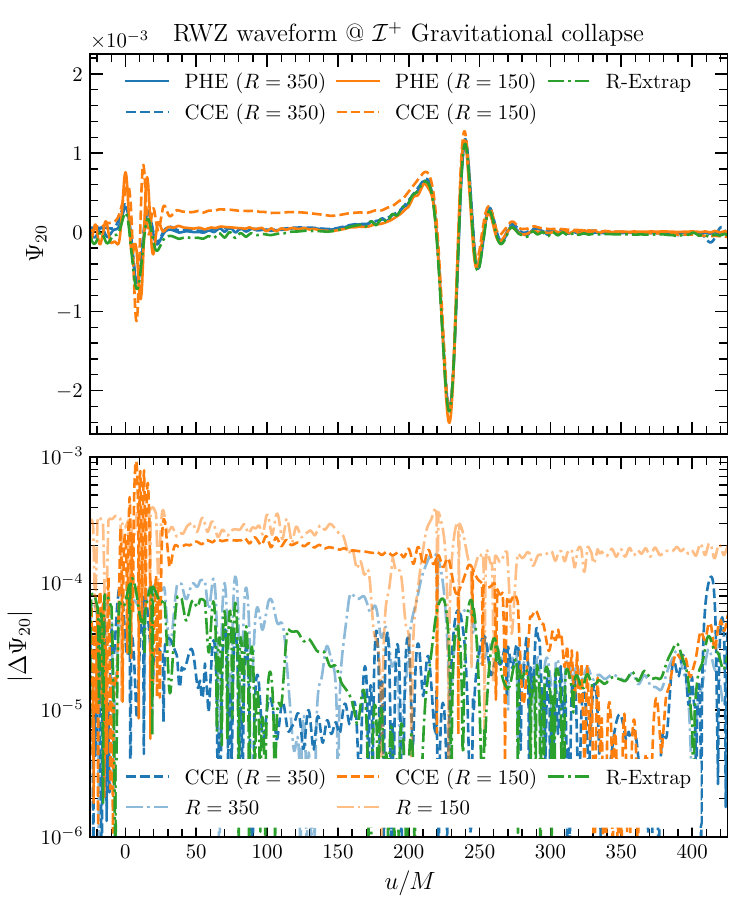}
  \caption{RWZ waveform at $\scri$ from the gravitational collapse of
    a rotating neutron star simulation.
    Top: $\Psi_{20}$ extrapolated with PHE (solid) and CCE (dashed)
    from worldtube data at $R=350\Mo$ (blue) and $150\Mo$ (orange),
    and best finite radius extrapolation (dashed-dotted green, at
    $R=150\Mo$). 
    Bottom: absolute differences between the PHE waveform and others: 
    $|\Delta\Psi_{20}|:=|\Psi^{{\rm PHE}}_{20}-\Psi^{X}_{20}|$, where
    the subtrahend $X$ waveforms are indicated in the legend.
    Light dashed-dotted lines refer to differences to RWZ waveforms
    extracted at at $R=350\Mo$ (blue) and $150\Mo$ (orange). These RWZ
    waveforms are not shown in the top panel.
  }
  \label{fig:rotcol}
\end{figure}

Figure~\ref{fig:rotcol} compares the dominant RWZ $\lm=20$ mode at
$\scri$ from different propagation (extrapolation) methods.
The top panel shows the RWZ waveforms: 
PHE from worldtubes at $R=350,150$ (blue and orange solid lines),
CCE from worldtubes at $R=350,150$ (blue and orange dashed lines)
and a ``best'' R-extrapolated using $R=150$ (green dashed-dotted line).
All waveforms have the characteristic \emph{precursor-burst-ringdown}
morphology~\cite{Davis:1972ud} and are broadly compatible with each other.
The bottom panel shows absolute differences
$|\Delta\Psi_{20}|:=|\Psi^{{\rm PHE}}_{20}-\Psi^{X}_{20}|$ 
between the PHE waveform and the others relative to the same
worldtube;
the legend indicates the subtrahend $X$ waveform. The bottom panel also shows in lighter
colors the differences with respect the two waveforms extracted at
$R=350$ (blue dashed-dotted) and $R=150$ (orange dashed-dotted).

The differences between PHE and CCE data converge to zero, indicating
that the waveforms computed with the two methods converge as the
worldtubes are placed at larger $R$. PHE-CCE differences in the
precursor for the choice $R=150$ are of the same order as the
differences between PHE and RWZ at $R=150$. The CCE extraction at
$R=150$ is therefore not sufficiently accurate due to the choice of
the worldtube. 
The differences PHE-CCE reduce 
of about an order of magnitude for the choice $R=350$. Some
unphysical (and small) oscillatory features in the precursor of the RWZ
waveforms at finite radius are propagated to $\scri$ by the
PHE. These features are reduced for worldtubes with larger radii, 
and are not present in the CCE waveforms.
The $R$-extrapolated waveform agrees well with both PHE and CCE.

\subsection{Binary black hole mergers}

Next we consider three waveforms from nonspining equal-mass binary
black hole simulations with very different morphology and generated in
a circular two-to-three orbits merger, a dynamical capture and a scattering
simulation. 
\subsubsection{Circular merger}

\begin{figure}[t]
  \centering
  \includegraphics[width=.46\textwidth]{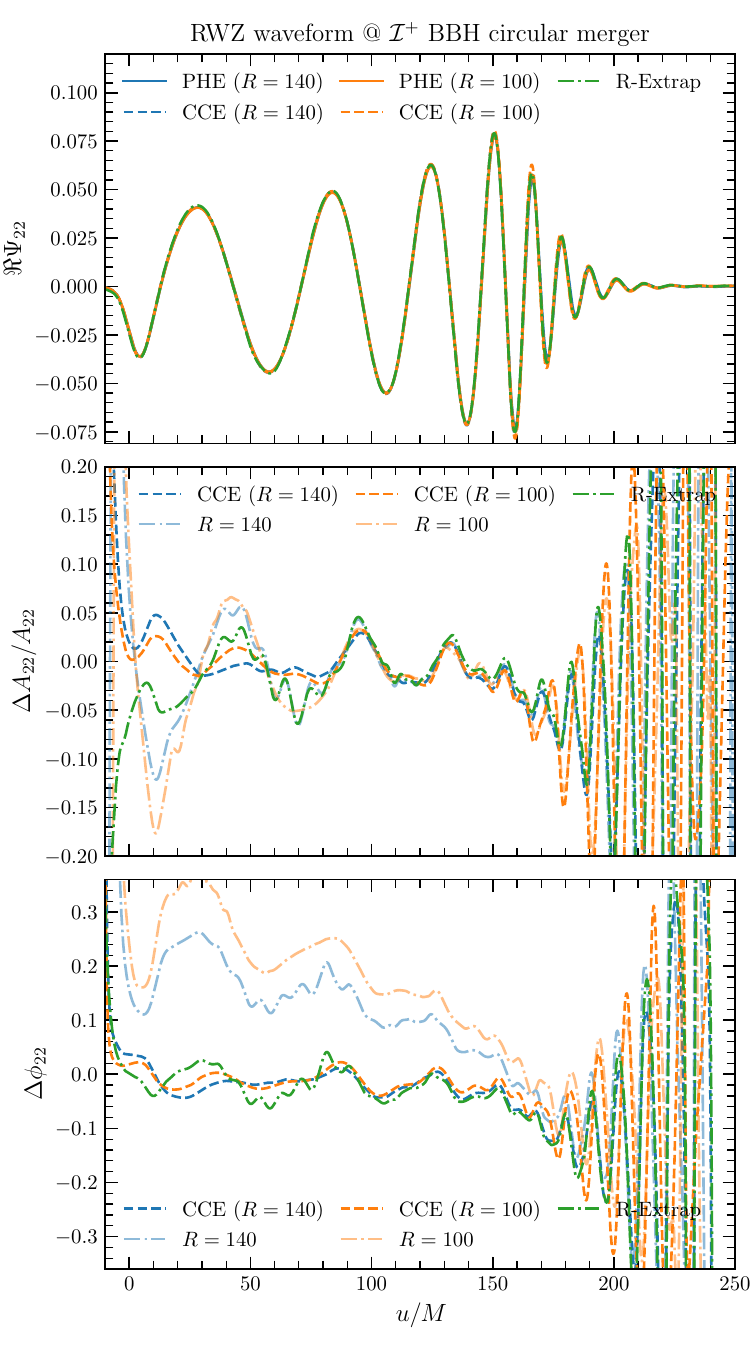}
  \caption{RWZ waveform at $\scri$ from a nonspinning equal-mass
    binary black hole circular merger.
    Top: Real part of $\Psi_{22}$ extrapolated with PHE (solid) and CCE (dashed)
    from worldtube data at $R=140\Mo$ (blue) and $100\Mo$ (orange),
    and best finite radius extrapolation (dashed-dotted green, $R=140\Mo$). 
    Middle: Relative amplitude differences between
    the PHE waveform and others: 
    $\Delta A_{22}/A_{22}:=(A^{{\rm PHE}}_{22}-A^{X}_{22})/A^{{\rm PHE}}_{22}$, where
    the subtrahend $X$ waveforms are indicated in the legend.
    Bottom: Phase differences between the PHE waveform and others: 
    $\Delta\phi_{22}:=\phi^{{\rm PHE}}_{22}-\phi^{X}_{22}$, where
    the subtrahend $X$ waveforms are indicated in the legend.
    Light dashed-dotted lines refer to differences to RWZ waveforms
    extracted at at $R=140\Mo$ (blue) and $100\Mo$ (orange). These RWZ
    waveforms are not shown in the top panel.
  }
  \label{fig:bbhcir}
\end{figure}

\begin{figure}[t]
  \centering
  \includegraphics[width=.49\textwidth]{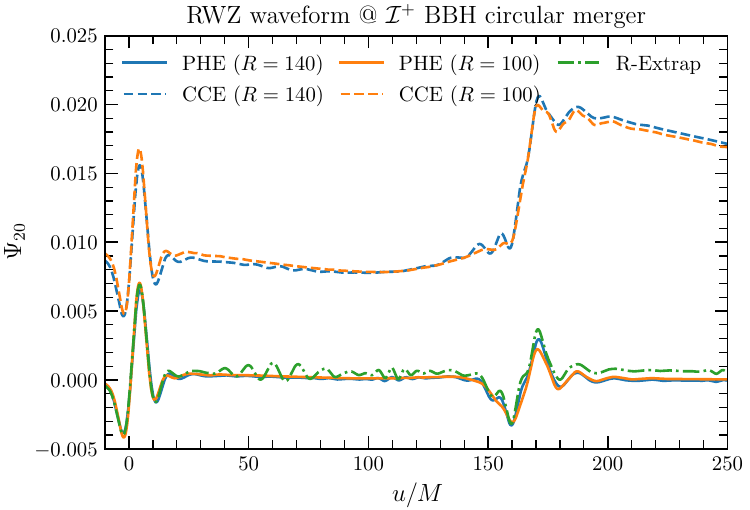}
  \caption{RWZ waveform at $\scri$ from a nonspinning equal-mass
    binary black hole circular merger.
    Real part of $\Psi_{20}$ extrapolated with PHE (solid) and CCE (dashed)
    from radii $R=140,100\Mo$ and best finite radius extrapolation
    (dashed-dotted green, $R=140\Mo$).     
    Nonlinear memory is captured only by the PHE linear propagation to $\scri$.
    The unphysical drift in the late ringdown of CCE is an artifact of
    the short-duration simulation.}
    \label{fig:bbhcir:20}
\end{figure}

The binary mass is $M\simeq1$ and the initial separation is such that
the binary undergoes about two orbit to merger, see
\eg~\cite{Brugmann:2008zz,Daszuta:2021ecf} for similar simulations.
Gravitational waves are extracted at coordinate (isotropic) radii 
$R=400,220,100,140$. Waveforms from the largest extraction radii are
significantly affected by the resolution of the Cartesian grid and we
thus focus on the smallest two radii.
The waveforms $\Psi^{\rm NR}(t)$ extracted at finite radius are
injected in a set of PHE simulations using
Eq.~\eqref{eq:gNR} and propagated to $\scri$.
The PHE grid is resolved with $N=201,401,801,1601,3201$ points. PHE
waveforms show convergence similarly to what discussed above; only the
highest resolution is discussed below.

Figure~\ref{fig:bbhcir} compares the dominant RWZ $\lm=22$ mode at
$\scri$ from different propagation (extrapolation) methods.
The top panel shows the RWZ waveforms: 
PHE from worldtubes at $R=140,100$ (blue and orange solid lines),
CCE from worldtubes at $R=140,100$ (blue and orange dashed lines)
and a ``best'' R-extrapolated using $R=140$ (green dashed-dotted line).
All waveforms have the characteristic \emph{inspiral-merger-ringdown}
morphology.
The bottom panels shows the relative amplitude differences
$\Delta A_{22}/A_{22}:=(A^{{\rm
    PHE}}_{22}-A^{X}_{22})/A^{{\rm PHE}}_{22}$
and the phase differences $\Delta\phi_{22}:=\phi^{\rm PHE}_{22}-\phi^{X}_{22}$;
the legend indicates the subtrahend $X$ waveform. The bottom panel also shows in lighter
colors the differences with respect the two waveforms extracted at
$R=140$ (blue dashed-dotted) and $R=100$ (orange dashed-dotted).

Amplitude relative differences PHE-CCE up to merger are below
${\lesssim}0.5$~\% and are compatible with the truncation error of 
  NR data at the considered resolution. In the late ringdown the
  differences increase significantly but this mostly due to the
  amplitude approaching zero and small relative dephasing of waveforms around
  their zeros.
The $R$-extrapolated waveforms well agree
  with the $\scri$ extrapolated waveforms. Finite radii extracted RWZ
  multipoles show differences of ${\lesssim}20$~\% in the early cycles
  but better agree at later times to merger.
  Phase differences PHE-CCE and PHE-$R$-extrapolated are ${\lesssim}\pm0.05$~rad,
  indicating a good agreement up to merger and early ringdown.
  Finite radius RWZ waveforms have instead differences as larger as
  ${\sim}0.2-0.3$~rad at early times over the two-orbits evolution.
  This is in agreement with previous work which indicated
  finite-extraction effect are more significant at early simulation
  times, see also below.

Figure~\ref{fig:bbhcir:20} compares the RWZ $\lm=20$ mode at
$\scri$ from different propagation (extrapolation) methods.
This mode has nonlinear memory
\cite{Christodoulou:1991cr,Blanchet:1992br} that can be captured only
by means of a nonlinear propagation to $\scri$. Consequently, neither
PHE nor extrapolation can capture the correct morphology of the
signal.
We stress that the CCE signals shown in the figure are strongly
affected by short-duration of the considered simulation. The CCE waveforms in the
late ringdown are unphysically drifting to lower value instead of
being constant. This is not a drawback of CCE, but simply a
consequence that 
the (integral) flux of gravitational waves is not entire captured by
the simulation considered here, \cf~\cite{Pollney:2010hs}.

\subsubsection{Dynamical capture}

\begin{figure}[t]
  \centering
  \includegraphics[width=.45\textwidth]{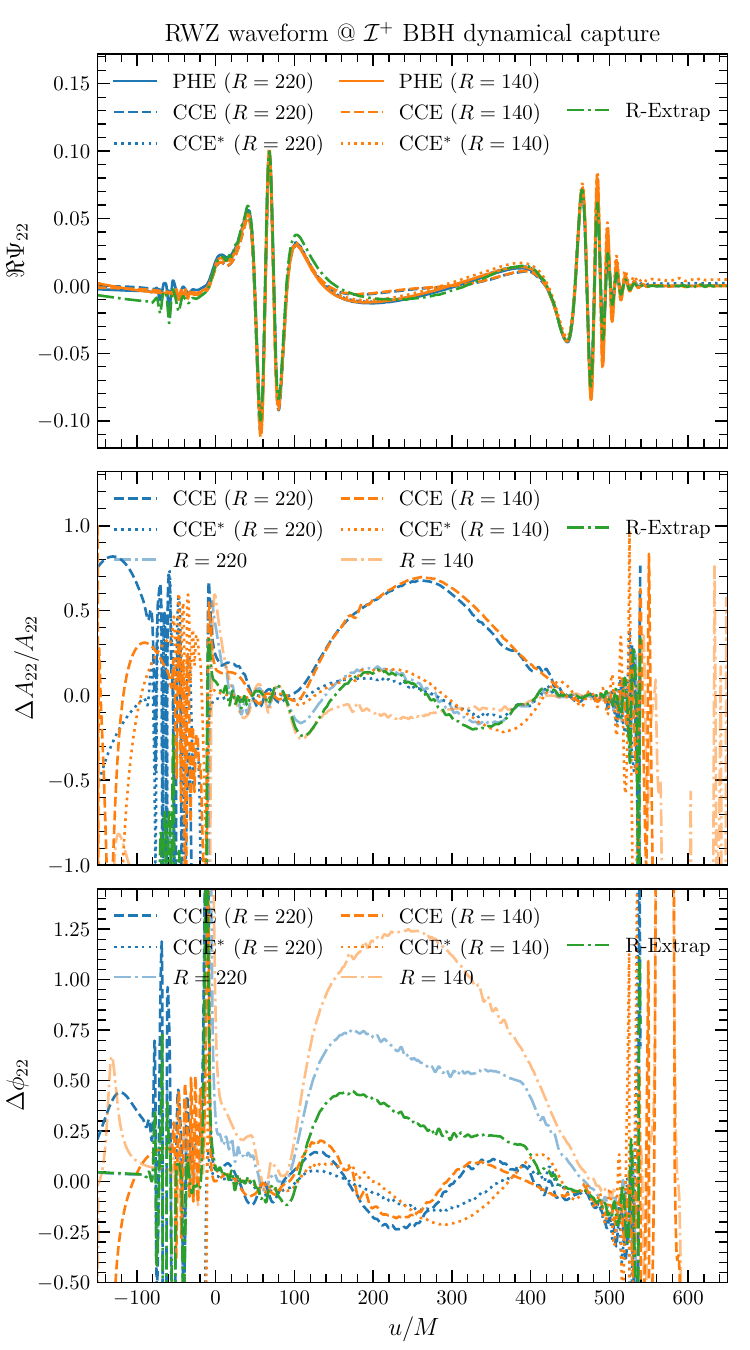}
  \caption{RWZ waveform at $\scri$ from a nonspinning equal-mass
    black hole dynamical encounter.
    Top: Real part of $\Psi_{22}$ extrapolated with PHE (solid),
    \pittnull~CCE (dashed) and \spectre~CCE (CCE$^*$, dotted)    
    from worldtube data at $R=140\Mo$ (blue) and $100\Mo$ (orange),
    and best finite radius extrapolation (dashed-dotted green, $R=220\Mo$). 
    Middle: Relative amplitude differences between
    the PHE waveform and others: 
    $\Delta A_{22}/A_{22}:=(A^{{\rm PHE}}_{22}-A^{X}_{22})/A^{{\rm PHE}}_{22}$, where
    the subtrahend $X$ waveforms are indicated in the legend.
    Bottom: Phase differences between the PHE waveform and others: 
    $\Delta\phi_{22}:=\phi^{{\rm PHE}}_{22}-\phi^{X}_{22}$, where
    the subtrahend $X$ waveforms are indicated in the legend.
    Light dashed-dotted lines refer to differences to RWZ waveforms
    extracted at at $R=220\Mo$ (blue) and $140\Mo$ (orange). These RWZ
    waveforms are not shown in the top panel.
  }
  \label{fig:bbhdyn}
\end{figure}

A different waveform morphology is given by the dynamical capture of
two black holes with total mass $M=1$~\cite{Fontbute:2025ixd}.
The punctures evolve in a close encounter, they separate and then
come together again to merge. Correspondingly, the waveform's
amplitude is characterized by two peaks connected by a lower frequency
signal and the final ringdown.
PHE data are computed at the same resolutions as for the circular
merger and we focus on the highest resolution $N=3201$.

Figure~\ref{fig:bbhdyn} compares the dominant RWZ $\lm=22$ mode at
$\scri$ from different propagation (extrapolation) methods.
PHE (from worldtubes at $R=220,140$, blue and orange solid lines),
CCE (from worldtubes at $R=200,140$, blue and orange dashed lines),
CCE obtained with \spectre~(from worldtubes at $R=220,140$, blue and orange dotted lines)
and the $R$-extrapolated (from $R=220$, green dashed-dotted line).
After computing CCE \spectre~data we apply a Bondi-van der Burg-Metzner-Sachs (BMS) transformation 
and map the strain to the superrest frame~\cite{Boyle:2015nqa}.

The figure illustrates that all waveforms have qualitatively similar
morphology with the double-peaked amplitude, the 
low-frequency signal and the ringdown.
The largest differences are in the low-frequency signal between the
peaks and in the late ringdown. 

Relative amplitude differences PHE-CCE show a good agreement during
the encounter and merger, while the transient before merger is not
well captured by the \pittnull~CCE extraction. The maximum difference
reaches up to $\Delta A_{22}/A_{22}\sim0.7$. This feature is ue to the
integration of the News function to the strain (or RWZ).
Indeed, the \spectre~CCE waveform shows significantly better
  performances here and agrees to the PHE with maximal amplitude
  differences of $\Delta A_{22}/A_{22}\sim 0.1$.

Phase differences PHE-CEE remain relatively flat with deviations of
order ${\lesssim}0.25$~rad, which is comparable to truncation errors of the NR
simulation.
Finite extraction and $R$-extrapolated RWZ waveform suffer of 
larger phase uncertainties up to ${\sim}1.25$~rad and
${\sim}0.75$~rad respectively accumulated in between the two amplitude
peaks. This highlights the need of computing waveform at $\scri$ in
order to accurately capture the entire signal for waveform modeling
purposes~\cite{Albanesi:2024xus}. 

\subsubsection{Scattering}

\begin{figure}[t]
  \centering
  \includegraphics[width=.49\textwidth]{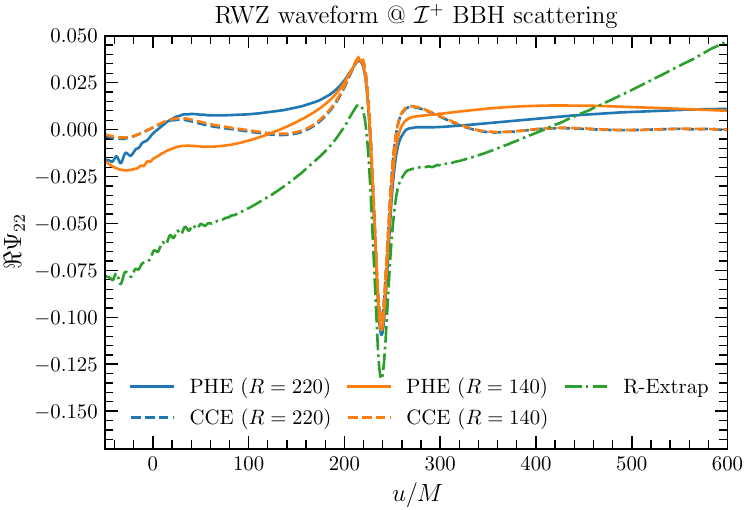}
  \caption{RWZ waveform at $\scri$ from a nonspinning equal-mass
    black hole scattering process.
    Real part of $\Psi_{22}$ extrapolated with PHE (solid) and CCE (dashed)
    from worldtube data at $R=200\Mo$ (blue) and $140\Mo$ (orange),
    and best finite radius extrapolation (dashed-dotted green, $R=220\Mo$).}
  \label{fig:bbhsca}
\end{figure}

We further consider a scattering problem with two nonspinning black
holes of total mass $M=1$ and initial energy (angular momentum) of
$E_{\rm in}/M\simeq1.023$ ($J_{\rm in}/M^2\simeq1.26$).
The initial data setup follow closely \cite{Damour:2014afa} with an
initial separation $d/M=100$.
Wave extraction in this type of simulations is challenging
because the punctures start at large separation and return quickly to 
the wave zone of the computational domain. We probe this scenario with a
simulation using the same grid setup as the circular
and dynamical capture \cite{Fontbute:2025ixd}.
PHE data are computed at the same resolution as for the circular
merger and we focus on the highest resolution $N=3201$.

Figure~\ref{fig:bbhsca} shows the real part of the $\lm=22$ mode at
$\scri$ from different propagation (extrapolation) methods.
The $R$-extrapolated waveform (dashed-dotted green) carries the
typical systematics of finite radius extraction: an unphysical drift
is visible in both the precursor and the
late tail and it is due to the close extraction radii.
These features are significantly amplified in waveforms obtained at finite extraction
radii (not shown). They cannot be entirely removed by considering
larger extraction radii since those NR waveform are additionally
affected by the progressively lower resolution of the Cartesian boxes.
Interestingly, the PHE removes most of the drift and delivers waveforms in
agreement with CCE around the amplitude's peak, \ie~the portion of
waveform contributing to the largest luminosity.

The \pittnull~CCE waveform is also affected by systematics
due to the reconstruction of the strain from the News
function.
Here the integration is performed with the fixed-frequency integration
algorithm~\cite{Reisswig:2010di} with a cutting frequency of
$f_0=0.007$, which appears close to optimal after manual experiments.
As discussed in detail
in the Appendix of \citet{Albanesi:2024xus}, the reconstruction is
ambiguous due to the difficulty of identifiying a cutting frequency
for the fixed-frequency integration algorithm (or a suitable
interval for the polynomial drift in the time-domain integral.)
As a consequence, the reconstruction procedure introduces systematics
that are more pronouced in the amplitude of the precursor and of the
tail. 
We have also checked \spectre~CCE data (not shown) and found a
large unphysical overall drift that cannot be straightforwardly cured by a
BMS frame rotation.
In summary, an improved NR setup with a more extended wave zone and higher 
resolutions than those considered here are required to mitigate 
waveform extraction effects in scattering simulations.

\subsection{Binary neutron star merger}

\begin{figure}[t]
  \centering
  \includegraphics[width=.46\textwidth]{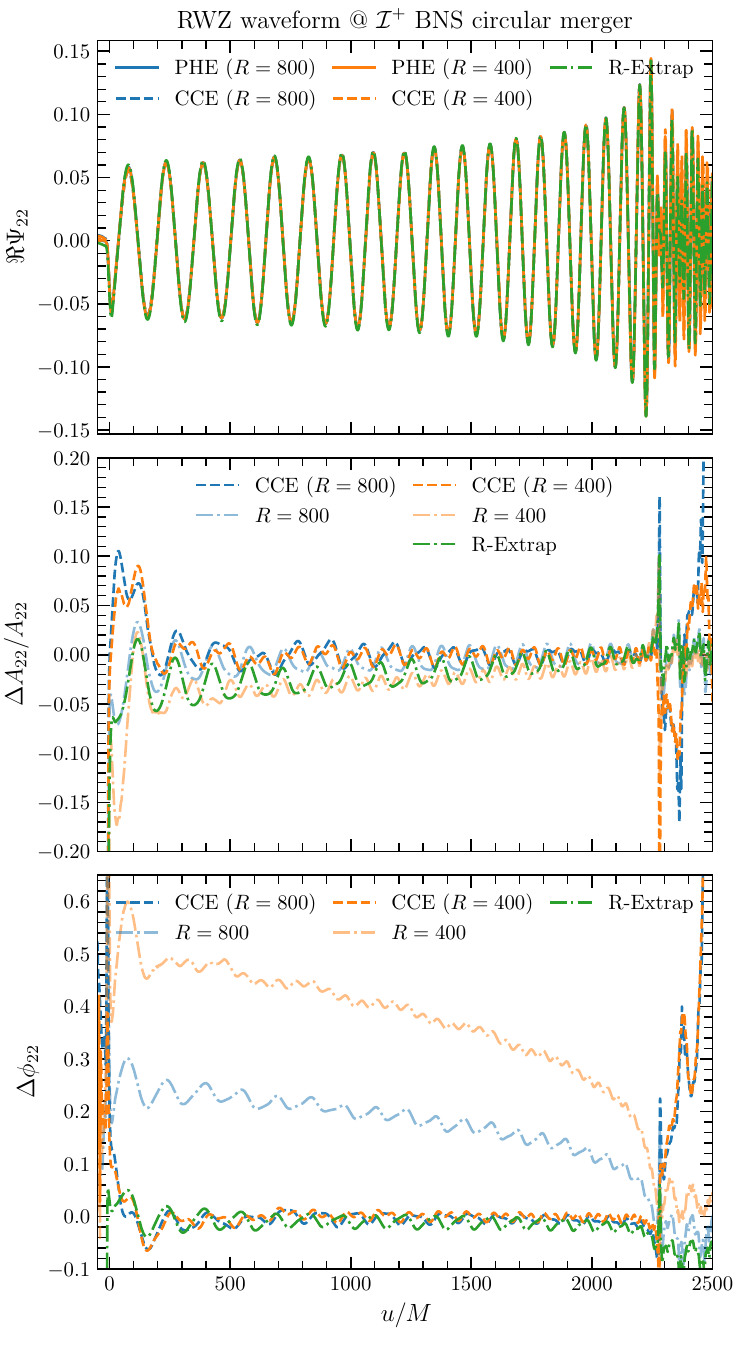}
  \caption{RWZ waveform at $\scri$ from a nonspinning equal-mass
    binary neutron star circular merger.
    Top: Real part of $\Psi_{22}$ extrapolated with PHE (solid) and CCE (dashed)
    from worldtube data at $R=800\Mo$ (blue) and $400\Mo$ (orange),
    and best finite radius extrapolation (dashed-dotted green, $R=800\Mo$). 
    Middle: Relative amplitude differences between
    the PHE waveform and others: 
    $\Delta A_{22}/A_{22}:=(A^{{\rm PHE}}_{22}-A^{X}_{22})/A^{{\rm PHE}}_{22}$, where
    the subtrahend $X$ waveforms are indicated in the legend.
    Bottom: Phase differences between the PHE waveform and others: 
    $\Delta\phi_{22}:=\phi^{{\rm PHE}}_{22}-\phi^{X}_{22}$, where
    the subtrahend $X$ waveforms are indicated in the legend.
    Light dashed-dotted lines refer to differences to RWZ waveforms
    extracted at at $R=800\Mo$ (blue) and $400\Mo$ (orange). These RWZ
    waveforms are not shown in the top panel.}
  \label{fig:bnscir}
\end{figure}

As a final application, we consider a circular equal-masses binary neutron star
merger. The binary mass is $M\simeq2.7$ and the initial separation is
such that the binary undergoes about
ten orbits to merger~\cite{Radice:2016gym,Dietrich:2017aum,Fontbute:2025ixd}.
Gravitational waves are extracted
at coordinate (isotropic) radii $R=800,600,400,200$.
PHE data are injected using Eq.~\eqref{eq:gNR} and we focus on the
highest resolution $N=3201$. 

Figure~\ref{fig:bnscir} shows the real part of the $\lm=22$ mode at
$\scri$ from different propagation (extrapolation) methods.
Relative amplitude differences PHE-CCE are essentially flat with
oscillations around ${\lesssim}5$~\% that reduce to ${\lesssim}1$~\%
to merger. The finite extraction RWZ and the $R$-extrapolated
waveforms 
are also consistent within these levels, thus providing a very good
representation of the waveform at $\scri$
Accumulated phase differences PHE-CCE (and PHE-$R$-extrapolated) to
merger are also flat and below ${\lesssim}0.05$~rad. These differences
are about an order magnitude smaller than those due to truncation
errors of the NR hydrodynamics. While the latter typically increase towards
merger, finite extraction phase error are of the
same order ${\sim}0.5$~rad at early times and reduces toward
merger~\cite{Bernuzzi:2011aq}.
This highlights the need for $\scri$ propagation or extrapolation in
high precision binary neutron star simulations~\cite{Bernuzzi:2016pie}.

Figure~\ref{fig:bnscir} also highlights significant differences in
both amplitude and phase in the postmerger waveform. This is mainly
due to the reconstruction of the strain from the News function in CCE
data. We stress that, in general, the postmerger signal is less 
accurate than the inspiral-merger signal and typically shows slower
convergence in the grid resolution due to the development of shocks in
the remnant. Therefore, these errors are also amplified by the strain
reconstruction procedure.

\section{Conclusion}
\label{sec:con}

PHE is a framework based on perturbative hyperboloidal evolutions 
for the propagation to $\scri$ of gravitational waveforms extracted at
finite radius in 3+1 numerical relativity.
The benchmark problems considered in this work include (3+1)D
simulations of gravitational collapse of rotating neutron stars,
binary black holes and binary neutror star mergers. They demonstrate
that PHE is a simpler and yet robust 
alternative approach to Cauchy-characteristic extraction for
astrophysical simulations in strong gravity.

Nonlinear effects in the waveform propagation are not captured by
PHE. Thus PHE cannot handle, for example, nonlinear tails and memory
effects,
\eg~\cite{Christodoulou:1991cr,Blanchet:1992br,Albanesi:2024fts,DeAmicis:2024eoy}. 
We plan to extend the linear PHE to second order to capture 
nonlinear waveform interactions, such as mode coupling and memory.  At second
order, the master equations acquire quadratic source terms built from products of
first-order RWZ modes, which drive the generation of new multipoles and
non-oscillatory offsets in the strain. The general covariant and
gauge-invariant formalism with 
explicit source constructions for Schwarzschild
perturbations is already developed~\cite{Brizuela:2006ne,Brizuela:2009qd,Spiers:2023mor},
see \eg~\cite{Ripley:2020xby,Bourg:2025lpd} for recent
applications. Implementing these techniques 
in a (1+1)D hyperboloidal code would 
allow us to propagate second-order corrections to $\scri$, delivering 
perturbative predictions of nonlinear memory, tail, and mode coupling in
astrophysically relevant gravitational-wave signals.

Similarly to CCE, PHE waveforms depend on the identification of an optimal 
radius for the worldtube. When Cartesian coordinates are employed in the 
3+1 simulation, the computational mesh is typically a hierarchy of Cartesian
boxes which progressively de-refine the wave zone. Therefore, the radius of the extraction
worldtube cannot be chosen arbitrarily large: resolutions effects can
severely affect the waveform quality.
Our work has shown that PHE waveforms robustly match CCE for
``optimal'' choices of the world tube. PHE can sometimes provide even
more robust prediction than CCE (see \eg~the rotational collapse
simulations) and reduce some of the unphysical features that are
present in waveforms extracted at finite radii.

RWZ waveforms extracted at finite radii are more prone to gauge
effects and numerical noise than Weyl $\psi_4$
waveforms~\cite{Fontbute:2025ixd}. Improved PHE waveforms at $\scri$ may be
obtained by propagating $\psi_{4}$ using Teukolsky equation.
Such variant of PHE can be straighfowardly implemented using the time-domain
hyperboloidal approach of \citet{Harms:2014dqa} in either (1+1)D or
(2+1)D. These $\psi_4$ PHE waveforms would still require double integration
to the strain and would thus carrying some of the systematics discussed in
this work, see also~\cite{Fontbute:2025ixd}.
Moreover, a quantitative investigation of the impact of
the (choice of the) background spin is required.

The PHE proposed here is another step towards the use of hyperboloidal
methods for the computation of gravitational radiation at null
infinity.
Future work may also consider a Cauchy-hyperboloidal extraction that incorporates nonlinearities in the waveform extraction by
leveraging on recent advances in numerical relativity with
hyperboloidal foliations.

\begin{acknowledgments}
  The authors thank David Hilditch for discussions and suggestions
  during the early stage of the project.
SB and JF aknowledge support by the EU Horizon under ERC
Consolidator Grant, no. InspiReM-101043372.
SA and SB acknowledge support from the Deutsche
Forschungsgemeinschaft (DFG) project ``GROOVHY'' (BE 6301/5-1
Projektnummer: 523180871). 
AZ acknowledges support by the National Science Foundation under
Grant No. 2309084.
Simulations were performed on SuperMUC-NG at the Leibniz-Rechenzentrum
(LRZ) Munich and and on the national HPE Apollo Hawk at the High
Performance Computing Center Stuttgart (HLRS). The authors acknowledge
the Gauss Centre for Supercomputing e.V. (\url{www.gauss-centre.eu})
for funding this project by providing computing time on the GCS
Supercomputer SuperMUC-NG at LRZ (allocations {\tt pn76li}, {\tt
  pn36jo} and {\tt pn68wi}). The authors acknowledge HLRS for funding
this project by providing access to the supercomputer HPE Apollo Hawk
under the grant number INTRHYGUE/44215 and MAGNETIST/44288. 
Postprocessing and development runs were performed on the ARA cluster
at Friedrich Schiller University Jena. The ARA cluster is funded in
part by DFG grants INST 275/334-1 FUGG and INST 275/363-1 FUGG, and
ERC Starting Grant, grant agreement no. BinGraSp-714626. 
\end{acknowledgments}

\end{document}